# Prévention des escarres chez les paraplégiques : une nouvelle approche par électrostimulation linguale.

A. Moreau-Gaudry[1,3], A. Prince[2], J. Demongeot[1,3], Y. Payan[3]

[1]Centre Hospitalier et Universitaire, Grenoble
[2]Centre Médico Universitaire Daniel Douady, St Hilaire du Touvet
[3]Laboratoire TIMC-IMAG-CNRS, Equipe GMCAO, Grenoble

*Résumé*-Cet article présente la faisabilité de l'utilisation par des sujets sains d'un nouveau dispositif de prévention des escarres chez les blessés médullaires paraplégiques. Cette prévention repose sur le principe de la substitution sensorielle qui consiste à suppléer à une modalité sensorielle déficiente (la sensibilité fessière du paraplégique), une modalité sensorielle fonctionnelle (sa sensibilité linguale). Le dispositif utilisé dans cette étude est constitué de trois parties : une matrice de 36 (6*6) électrodes placées sur la face supérieure de la langue; un ensemble de 144 (12*12) capteurs de pression disposés dans un coussin sur lequel le sujet s'assoit; un ordinateur portable qui, connecté de manière filaire au coussin et à la matrice, réalise une interface entre les deux dispositifs précédents. L'expérimentation a consisté à activer, par électrostimulation, les récepteurs tactiles de la langue selon une « image directionnelle » et à enregistrer les variations de pression au niveau du coussin, le patient devant modifier sa posture selon l'information de direction décodée. Les résultats obtenus montrent, chez 10 sujets sains, dans 92% ± 7.9% des cas, une réponse en pression adaptée à l'image directionnelle électrostimulée. Ces probants premiers résultats permettent d'envisager l'utilisation d'un tel dispositif à la prévention des escarres de pression chez les sujets paraplégiques : suite à l'identification des zones potentielles de souffrance tissulaire (zones de surpression) à partir d'une analyse automatique par l'ordinateur de la carte de pression, l'électrostimulation linguale de la « meilleure » direction de mobilisation posturale déduite de cette analyse permettrait au patient paraplégique de modifier sa posture et ainsi de s'affranchir de cette souffrance par homogénéisation des pressions d'appui. En outre, cet article rapporte les améliorations technologiques et ergonomiques réalisées depuis l'utilisation de ce premier dispositif en vue de son intégration dans le quotidien du patient.

## I. Introduction

### A. Escarre & Paraplégie

Une escarre est définie comme une « détérioration localisée de la peau et du tissu sous jacent consécutive à la pression, à la friction, au cisaillement ou à une combinaison de ces différents facteurs » [1]. Privés de leurs capacités sensorielles et motrices, les lésés médullaires paraplégiques, assis dans leur fauteuil, ne possèdent plus le réflexe inconscient de changer de position comme le fait une personne valide assise. Ils peuvent ainsi développer des escarres, aux conséquences parfois dramatiques. Chez le paraplégique, elles sont principalement localisées au niveau des régions ischiatique, sacrée et trochantérienne [1]. En effet, une pression de 60 à 80 mmHg, maintenue pendant 2 à 3 heures, peut provoquer des escarres [2,3].

### B. Solutions actuelles

Le contrôle de la charge tissulaire indispensable à la prévention des escarres a engendré le développement de coussins et de matelas "anti-escarre". Ces matériels permettent, par diminution des pressions à l'interface fesses-coussins, de diminuer les forces qui agissent à l'interface muscle-ischion [4]. L'objectif de ces matelas/coussins est ainsi d'obtenir, par redistribution des forces de pression, une pression suffisamment faible pour ne pas provoquer de plaies de pression. Ces matelas réducteurs de pression, constitués de gel, de mousse, d'eau ou d'air, font partie des stratégies actuelles de prévention des plaies dans les milieux hospitaliers. Toutefois, ces supports ne suffisent pas et les techniques de transfert de poids constituent toujours le meilleur moyen de prévention. Soulèvements, changements de position, modifications des appuis sont à réaliser régulièrement afin d'assurer une bonne répartition pressionnelle et ainsi prévenir la formation de cette lésion [5]. Des dispositifs récemment développés détectent, au moyen de capteurs de pression positionnés sous les fessiers, les régions en surpression trop longuement exposées. Ils renvoient cette information au patient et au médecin sous forme d'une carte de couleur affichée sur écran ou plus simplement par un moyen externe (signal sonore ou lumineux par exemple). Les zones à risque sont ainsi identifiables. Néanmoins, ce dispositif présente deux inconvénients majeurs: il limite la mobilité du patient et il le contraint à être attentif en permanence à l'écran de contrôle (au signal sonore ou au signal lumineux). Un des objectifs de cette recherche est de s'affranchir de ces inconvénients par la mise en œuvre du principe de la substitution sensorielle.

*C. Principe de la substitution sensorielle*

Au cours de leurs recherches, Bach-y-Rita et ses collaborateurs [6,7] ont montré que des stimuli caractéristiques d'une modalité sensorielle (la vision) pouvaient être remplacés par des stimuli d'une autre modalité sensorielle (le touché). Le premier développement d'un système de substitution visuo-tactile a été réalisé dans le but de fournir une information visuelle distale aux aveugles [8]. Ce système, le « Tactile Vision Substitution System » (TVSS) était composé d'une matrice de 400 électrodes tactiles (20*20) placées sur différentes parties du corps comme le front ou l'abdomen et permettait la traduction, en "images tactiles", d'images capturées par une caméra. Bach-y-Rita et ses collaborateurs ont montré, grâce à ce système, que les aveugles étaient capables de reconnaître des formes du monde extérieur à travers la modalité tactile [9]. Après un temps d'adaptation, le sujet aveugle oublie les stimulations sur la peau et perçoit les objets comme étant devant lui. Après plusieurs semaines d'apprentissage, des potentiels évoqués étaient observés au niveau de l'aire visuelle jamais activée auparavant. Autrement dit, les aveugles, sujets aux expérimentations de substitution sensorielle, commençaient par percevoir l'électrostimulation comme une activité tactile classique, puis se mettaient véritablement à "voir" à travers la paume, les doigts ou le dos : « nous ne voyons pas avec les yeux » [9].

*D. Choix d'une modalité sensorielle de substitution*

Dans le cadre de cette recherche, nous nous sommes orientés vers la langue, comme organe de substitution sensorielle, pour deux raisons. La première est en rapport avec les travaux de Bach-y-Rita sur la substitution sensorielle, au cours desquels il a récemment convergé vers cet organe [10]. En effet, la langue, organe du corps humain possédant la plus grande densité de récepteurs tactiles, dispose d'une capacité discriminative supérieure à celle du bout des doigts [11]. De plus, la forte conductivité de la salive permet la mise en œuvre d'une technologie basée sur une électro-stimulation à très faible énergie. La seconde est la conséquence directe des traumatismes médullaires : alors que la plupart des centres nerveux sensitifs et moteurs rachidiens peuvent être lésés chez les traumatisés rachidiens, la langue se trouve souvent être une structure préservée chez ces patients paraplégiques.

*E. Objectif*

Notre projet de recherche vise à compenser les déficits sensoriels des paraplégiques par l'utilisation d'un dispositif d'électrostimulation linguale (ensemble d'électrodes placées en bouche) couplé à des capteurs de pression inclus au sein d'un coussin de pression, dans le but de réduire l'incidence des escarres de pression chez le paraplégique. Dans cette étude préliminaire, nous nous sommes intéressés à la faisabilité d'une telle approche chez des sujets sans lésion médullaire.

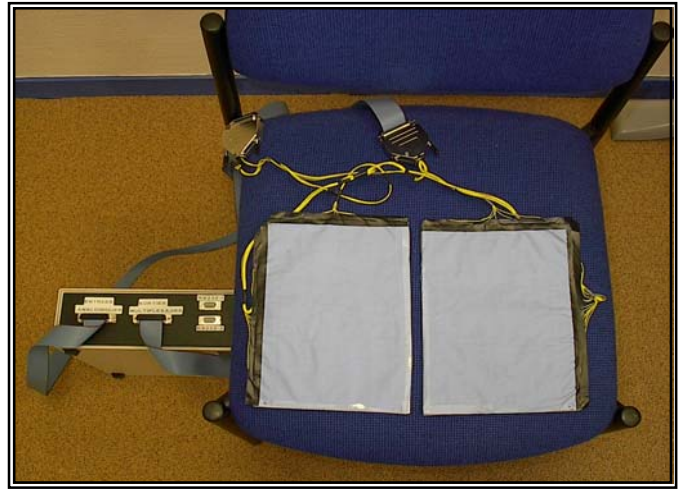

Fig. 1. Le coussin de pression (prototype).

## II. MATERIELS & METHODES

*A. Le coussin de pression*

Nous avons développé, en collaboration avec la société Léas®, un coussin de pression (prototype) qui permet l'acquisition, en temps réel (f=10Hz), des pressions appliquées sur les fessiers du paraplégique lorsque ce dernier est assis dessus. Ce coussin est constitué de deux hémi-coussins droit et gauche (Fig. 1). Chaque hémi-coussin comporte 72 (6x12) capteurs de pression. Chaque capteur de pression a une surface d'environ 1cm². La répartition des capteurs dans le coussin est non linéaire afin de permettre une meilleure résolution dans les zones ischiatiques où les risques d'escarres sont plus élevés. Ces capteurs fonctionnent grâce à une poudre semi-conductrice répartie uniformément dans une enveloppe de polymère. Cette poudre possède des propriétés élastiques et agit selon le principe de percolation : lorsque le volume du capteur change sous l'effet d'une force de pression, sa conductivité augmente et la résistance au sein du capteur baisse. La variation de courant envoyé à travers la poudre est ainsi fonction de la pression exercée. Chaque capteur est relié à un système électronique qui permet la mesure de son potentiel électrique codé sur 4 bits (16 niveaux de pression par capteur).

*B. Le TDU*

Un prototype, nommé « Tongue Display Unit » (TDU) a été développé et validé sur une population de non voyants pour lesquels les informations visuelles captées par caméra vidéo étaient transcodées en stimulations électrotactiles linguales [12]. Dans cette étude, nous utilisons une version simplifiée de ce prototype, constituée d'une matrice de 36 électrodes tactiles (6×6) collées sur une bande plastique et connectées à un système électronique externe (Fig. 2). Le sujet peut maintenir les électrodes au contact de sa langue, la bouche fermée. La salive offrant une bonne conductivité, le TDU requiert seulement un voltage de sortie de 5 à 15V et un courant de 0.4 à 4mA pour stimuler les récepteurs linguaux.

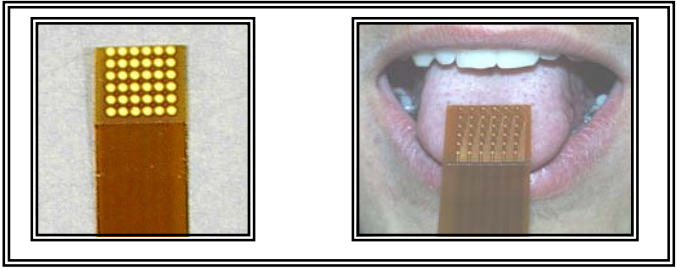

Fig. 2. Les 36 électrodes du TDU sont placées
sur et au contact de la langue.

Ainsi, lorsqu'une électrode est activée, le sujet ressent, à ce niveau, un « picotement » à la surface de sa langue. Dans la présente étude, nous avons maintenu constante la fréquence de stimulation (50 Hz), seule l'intensité de la stimulation variait en fonction de la sensibilité du sujet.

### C. La méthodologie

L'objectif est de montrer la capacité d'un sujet, sans lésion médullaire, à percevoir et interpréter l'information fournie par les électrodes du TDU placées sur la langue, puis à adopter une attitude posturale adaptée à l'information décodée. Dix sujets sains volontaires du laboratoire TIMC (5 de sexe féminin), après signature d'un consentement éclairé, d'âge moyen 26.2 ans (âge minimal : 24 ans, âge maximal : 37 ans) ont accepté de participer à une première validation clinique constituée de trois parties : la première a consisté à *calibrer* la stimulation électrique linguale, du fait de l'anisotropie spatiale individuelle de la sensibilité linguale. Pour chaque individu, cette calibration a été effectuée en modulant l'intensité de la stimulation linguale en fonction de la perception rapportée par le sujet. La seconde partie était consacrée à l'*apprentissage*. Il a été demandé au sujet d'identifier le groupe d'électrodes activées parmi 4 groupes. Ces groupes étaient constitués des 6 électrodes des bords respectivement antérieur, postérieur, droit et gauche de la matrice d'électrode, dans le référentiel du patient (Fig. 3). Enfin, la dernière partie regroupait 10 *épreuves cliniques*, chaque épreuve s'étant déroulée de la manière suivante : la carte de pression de chaque sujet assis sur le coussin de pression, au repos, a été acquise pendant une durée de 3 secondes. Une des 4 électrostimulations linguales a ensuite été envoyée sur le TDU. Après avoir identifié la région de la matrice activée, le sujet devait mobiliser son buste selon la direction indiquée par l'électrostimulation linguale (les stimulations antérieure, postérieure, gauche, droite correspondaient respectivement à une demande de déplacement du buste en avant, en arrière, à gauche, à droite). Après 3 secondes, temps accordé au sujet pour mobiliser son buste selon l'information directionnelle reçue sur la langue, la nouvelle carte de pression du sujet, assis, au repos, a été enregistrée. Un score unitaire ou nul a été attribué à chaque épreuve si et seulement si le déplacement des barycentres des cartes de pression, avant et après mouvement du buste, concordait ou non avec l'information linguale électro-stimulée. Ainsi, pour chaque individu, nous avons obtenu un score total côté sur 10, tous les sujets ayant réalisé les mêmes épreuves cliniques.

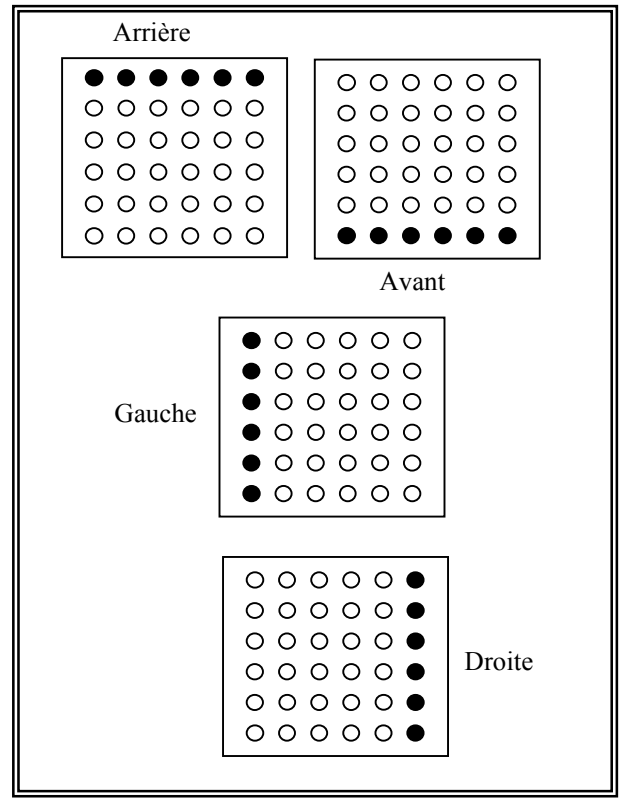

Fig. 3. Schéma des quatre stimulations envoyées sur la matrice des 36 électrodes du TDU pour indiquer la direction souhaitée de mobilisation du buste à adopter (en arrière, en avant, sur la gauche, sur la droite).

### III. RESULTATS

Chaque sujet, ayant accepté de participer à cette épreuve expérimentale, l'a mené à terme. L'étape de calibration individuelle n'a présenté aucune difficulté. L'apprentissage a été obtenu au bout de 4 à 8 stimulations. Les épreuves cliniques relatives au déplacement du buste selon l'information directionnelle électrostimulée se sont déroulées sans incident. Le score moyen obtenu était de 9.2 déplacements adaptés, en pression, à l'information directionnelle électrostimulée, avec un écart type de 0.79, un minimum à 8 et un maximum à 10.

### IV. DISCUSSION

Cette étude préliminaire montre, chez des patients sans lésion médullaire, la faisabilité et la fonctionnalité (92% de réussite) de la séquence « Stimulation linguale TDU - Evaluation adaptée de l'information électrostimulée - Réaction posturale adaptée en pression à l'information directionnelle électrostimulée ».

*A. Evaluation clinique future*

L'évolution naturelle de cette approche va consister à évaluer, sur une population de patients paraplégiques, la faisabilité non pas de la séquence mais de la boucle Stimulation / Evaluation / Réaction adaptée en pression. En effet, la modification de la posture (Réaction) adaptée en pression à l'information linguale directionnelle électrostimulée est à l'origine d'une nouvelle distribution spatiale des forces de pression. En l'absence de mouvement et de sensibilité fessière, cette nouvelle distribution peut-être à l'origine, à plus ou moins long terme, d'une nouvelle souffrance tissulaire par surpression locale. Un de nos objectifs est de programmer l'ordinateur de manière à ce qu'il identifie, par analyse automatique de la carte des pressions exercées à l'interface région fessière/coussin, les zones de souffrance tissulaire à risque d'escarre (zone de surpression). Ce premier objectif réalisé, l'ordinateur pourrait alors être programmé de manière à proposer, à partir de cette analyse, le changement de posture optimal (traduit sur la matrice TDU) qui soulagera au mieux ces zones de surpression. L'électrostimulation au niveau lingual de cette information directionnelle permettrait ainsi d'achever la séquence précédemment décrite puisque cette nouvelle information linguale sera à l'origine d'un déplacement postural, donc d'une nouvelle distribution des forces de pression et ainsi de suite. Afin de montrer la faisabilité de cette approche, une recherche biomédicale prospective, monocentrique, randomisée, contrôlée, en groupe parallèle, équilibrée, ouverte, devrait débuter au mois d'Avril 2006, avec 24 (12/12) sujets paraplégiques du Centre Medico Universitaire Daniel Douady (CMUDD) de Saint Hilaire du Touvet (38).

A terme, nous pensons que les sujets utilisant le système TDU parviendront à adopter, par réflexe inconscient, des mouvements de compensation de la même façon qu'une personne valide. Autrement dit, nous espérons une prise en compte de type « bas niveau » des mouvements à effectuer en fonction de l'information de posture transmise par le TDU en vue d'une normalisation des pressions à l'interface région fessière/coussin, et donc une prévention de la formation d'une escarre.

Plus généralement, nous espérons, par analogie avec les travaux de P. Bach-y-Rita sur la substitution sensorielle chez les aveugles, que les lésés médullaires pourront "ressentir" les pressions exercées sur cette région à partir de cartes de pression "affichées" sur leur langue, cartes de pression brutes ou prétraitées à partir desquelles ils adapteront leur posture en conséquence.

*B. Evolution matérielle en vue d'une intégration dans le quotidien du patient*

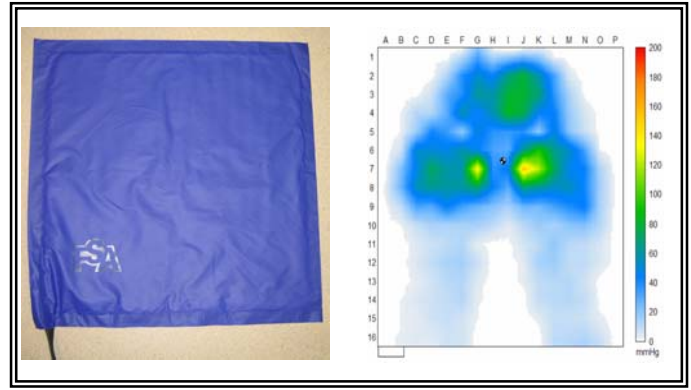

Fig. 4. Nappe de pression FSA de la Société Vista Medical

Une intégration dans le quotidien du patient ne peut-être obtenue que par amélioration de la fiabilité du matériel ainsi que de son ergonomie. En ce qui concerne les capteurs de pression, nous venons de signer un accord de collaboration avec la société Vista Medical, société qui développe une nappe de pression de 32x32 capteurs avec possibilité de calibration automatique pneumatique initiale (Fig. 4).

En ce qui concerne l'électrostimulation linguale, une approche filaire étant utopique pour une acceptation du dispositif par le patient, nous avons développé, en collaboration avec les sociétés CORONIS-SYSTEMS et GUGLIELMI TECHNOLOGIES DENTAIRES, une prothèse orthodontique contenant une version miniaturisée et sans fil du TDU (Fig. 5). Plus précisément, le kit TUD miniaturisé est inclus dans une prothèse palatine obtenue à partir des empreintes dentaires du patient, donc idéalement adaptée à l'anatomie du patient et ainsi garant d'une acceptabilité maximale. Ce kit communique par voie hertzienne avec l'ordinateur, lui-même connecté à la nappe de pression. Alors que ces difficultés ergonomiques étaient jusqu'à présent perçues comme un frein à la substitution sensorielle [12], nous espérons, par ces améliorations, développer un véritable dispositif de substitution sensorielle accepté par le patient dans son quotidien.

## V. Conclusion

Chez 10 sujets, sans lésion médullaire, la réponse en pression enregistrée au niveau de l'interface région fessière/fauteuil suite au déplacement du buste est adaptée à l'information directionnelle préalablement électrostimulée. Ces résultats permettent d'envisager l'utilisation de ce dispositif à la prévention des escarres de pression chez les sujets paraplégiques, par identification des zones de souffrance tissulaire à partir d'une analyse automatique par l'ordinateur de la carte de pression, et électrostimulation linguale d'une direction de mobilisation posturale à adopter par le patient paraplégique pour corriger cette surpression. De plus, l'amélioration de la fiabilité et de l'ergonomie du système développé pourrait permettre, si la faisabilité de la boucle Perception/Analyse/Action est montrée, chez les paraplégiques, un investissement naturel du quotidien du patient garant d'une compliance thérapeutique optimale.

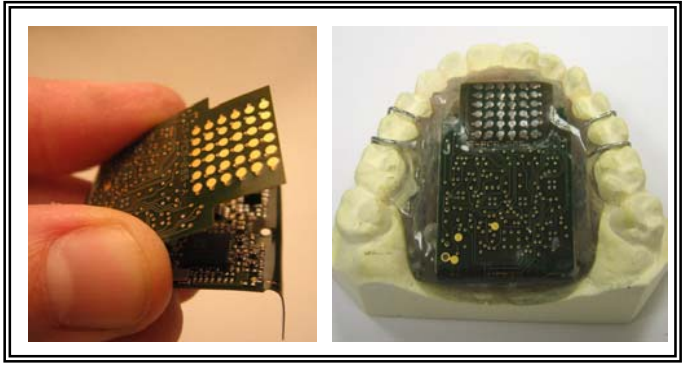

Fig. 5.  Prototype de TDU sans fil 6x6 (gauche) inclus dans une prothèse orthodontique (droite).

## REFERENCES

[1] "Pressure Ulcer Treatment Guidelines," retrieved January 10, 2006, from http://www.epuap.org/gltreatment.html.
[2] M. Kosiak, "A mechanical resting surface: its effects on pressure distribution", *Arch. Phys. Méd. Rehabil.*, vol. 57, no. 10, pp. 481-4, October 1976.
[3] J. Maklebust, "Pressure ulcers: etiology and prevention", *Nurs Clin North Amer*, vol. 22, no. 2, pp. 359-77, June 1987.
[4] JK. Hedrick-Thompson, "A review of pressure reduction device studies". *J Vasc Nurs*, vol. 10, no. 4, pp. 3-5, December 1992.
[5] RM. Letts, *Le positionnement : principes et pratique*, Montréal/Paris : Décarie; Maloine, 357 p., 1995
[6] P. Bach-y-Rita, C. Collins, F. Saunders, B. White and L. Scadden, "Vision Substitution by Tactile Image Projection", *Nature*, vol. 221, pp. 963-964, March 1969.
[7] P. Bach-y-Rita, "Late postacute neurological rehabilitation: neuroscience, engineering and clinical programs", *Archives of Physical Medicine and Rehabilitation*, vol. 84, no. 8, pp. 1100-8, August 2003.
[8] CC. Collins and P. Bach-y-Rita, "Transmission of Pictorial Information through the Skin," *Advan. Biol. Med. Phys.*, no. 14, pp. 285-315, 1973.
[9] Bach-y-Rita, *"Brain mechanisms in sensory substitution"*, New York: *Academic Press*, pp. 182, 1972
[10] P. Bach-y-Rita, K.A. Kaczmarek, M.E. Tyler and J. Garcia-Lara, "Form Perception with a 49-points Electro-Tactile Stimulus array on the Tongue: A Technical Note", *Journal of Rehabilitation Research and Development*, vol. 35, no. 4, pp. 427-430, October 1998.
[11] E. Sampaio, S. Maris, P. Bach-y-Rita, "Brain plasticity: 'visual' acuity of blind persons via the tongue", *Brain Research*, 908, pp. 204-207, 2001
[12] P. Bach-y-rita and SW. Kercel, "Sensory substitution and the human-machine interface," *Trends Cog. Sci.I,* vol. 7, no 12, December 2003.